\documentclass[a4paper,twoside,11pt]{article} 
\newlength{\defaultparindent}
\usepackage{latexsym}
\usepackage{amsmath}
\usepackage{amsfonts}
\usepackage{amsthm}
\usepackage{bbold}
\usepackage[applemac]{inputenc} 
\usepackage{multirow}

\usepackage[arXiv]{optional} 

\opt{AACA}{
\def\cal{\mathcal}
}

\opt{x,std,JMP,AACA}{
\input{mbh_defs_TeX}
}

\opt{arXiv}{
\newtheorem{MS_Proposition}{Proposition}
\newtheorem{MS_Corollary}[MS_Proposition]{Corollary}
\def\Cl{{\cal C}\ell}
\def\End{\textrm{End}}
\def\R{{\mathbb{R}}} 
\def\C{{\mathbb{C}}} 
\def\my_span#1{\mbox{Span}\left(#1\right)} 
\def\Identity{{\mathbb{1}}} 

\newcommand{\comm}[2]{\ensuremath{\left[ #1, #2 \right]}}
\newcommand{\anticomm}[2]{\ensuremath{\left\{ #1, #2 \right\}}} 
}

\opt{JMP}{
}

\def\h_eigen{\eta}
\def\g_eigen{\theta}

\begin{document}

\opt{x,std,arXiv,JMP}{
\title{\bf The Extended Fock Basis of Clifford Algebra}

\author{\\
	\bf Marco Budinich\\
	Dipartimento di Fisica\\
	Università di Trieste \& INFN\\
	Via Valerio 2, I - 34127 Trieste, Italy\\
	\texttt{mbh@ts.infn.it}\\
	\texttt{http://www.ts.infn.it/\~{ }mbh/MBHgeneral.html}\\
	\\
	Published in: {\em Advances in Applied Clifford Algebras}, 2011\\
	{\small DOI:10.1007/s00006-011-0316-2}
	}
\date{  }

\maketitle
}

\opt{AACA}{
\title[The EFB of Clifford Algebra]{The Extended Fock Basis of Clifford Algebra}

\author{Marco Budinich}
\address{Dipartimento di Fisica\\
	Università di Trieste \& INFN\\
	Via Valerio 2, I - 34127 Trieste, Italy}
\email{mbh@ts.infn.it}
}

\begin{abstract}
We investigate the properties of the Extended Fock Basis (EFB) of Clifford algebras \cite{BudinichM_2009} with which one can replace the traditional multivector expansion of $\Cl(g)$ with an expansion in terms of simple (also: pure) spinors. We show that a Clifford algebra with $2 m$ generators is the direct sum of $2^m$ spinor subspaces $S$ characterized as being {\em left} eigenvectors of $\Gamma$; furthermore we prove that the well known isomorphism between simple spinors and totally null planes holds {\em only} within one of these spinor subspaces.
We also show a new symmetry between spinor and vector spaces: similarly to a vector space of dimension $2m$ that contains totally null planes of maximal dimension $m$, also a spinor space of dimension $2^m$ contains ``totally simple planes'', subspaces made entirely of simple spinors, of maximal dimension $m$.
\end{abstract}

\opt{x,std,arXiv,JMP}{
\noindent{\bf Keywords:} {Clifford algebra, Spinors, mathematical physics, Fock basis.}
}

\opt{AACA}{
\keywords{Clifford algebra, Spinors, mathematical physics, Fock basis.}
\maketitle
}

\section{The extended Fock basis of Clifford algebra}
\label{Clifford_algebra_and_EFB}
We begin summarizing the main properties of the Extended Fock Basis (EFB) of Clifford algebra introduced in \cite{BudinichM_2009}. We will consider Clifford algebras (see e.g.\ \cite{Chevalley_1954}) with an even number of generators $\gamma_1, \gamma_2, \ldots, \gamma_{2 m}$ over field $F$. These are simple algebras of dimension $2^{2 m}$ and with vector space $F^{2 m} := V$. The results that follow hold both for $F = \C$ and $\R$ with signature
$$
\gamma_{2 i - 1}^2 = 1 \quad \gamma_{2 i}^2 = -1 \qquad i = 1,\ldots,m
$$
we leave to the reader the simple adjustments for the two cases. Given the $\R^{2 m}$ signature we indicate the Clifford algebra with $\Cl(m, m)$ that has been deeply studied also in \cite{Rodrigues_2007}. A Clifford algebra can be seen as the direct sum of its graded parts: field $F := F^{(0)}$, vectors $V := F^{(1)}$ and multivectors $F^{(k)}, \; 1 < k \le 2 m$
\begin{equation}
\label{Clifford_direct_sum}
\Cl(m, m) = F^{(0)} \oplus F^{(1)} \oplus \cdots \oplus F^{(2 m)}
\end{equation}
and is graded isomorphic to $F( 2^m )$, the algebra of matrices of size $2^m \times 2^m$.

The EFB essentially extends to the entire algebra the Fock basis~\cite{BudinichP_1989} of its spinorial part and renders explicit the construction $\Cl(m, m) \cong \overset{m}{\otimes} \Cl(1, 1)$ so that many properties of $\Cl(m, m)$ can be proved in $\Cl(1, 1)$. We start from the null, or Witt, basis of the vector space $V$ that takes the form:
\begin{equation}
\label{formula_Witt_basis}
p_{i} =\frac{1}{2} \left( \gamma_{2i-1} +\gamma_{2i} \right)
\quad \textrm{and} \quad
q_{i} =\frac{1}{2} \left( \gamma_{2i-1} -\gamma_{2i} \right)
\quad i = 1,2, \ldots, m
\end{equation}
that, with $\gamma_{i} \gamma_{j} = - \gamma_{j} \gamma_{i}$, easily gives (here $\anticomm{q_i}{p_j} := q_i p_j + p_j q_i$)
\begin{equation}
\label{formula_Witt_basis_properties}
\anticomm{p_{i}}{p_{j}} = \anticomm{q_{i}}{q_{j}} = 0
\qquad
\anticomm{p_{i}}{q_{j}} = \delta_{i j} \Identity
\end{equation}
that imply $p_i^2 = q_i^2 = 0$, at the origin of the name ``null'' (also: isotropic) given to these vectors.

\begin{table}
\label{piqi_table}
\centering
\begin{tabular}{c | c | c c c c |}
\multicolumn{2}{c}{} & \multicolumn{4}{c}{$\phi_i$} \\
\cline{3-6}
\multicolumn{1}{c}{} & & $q_i p_i$ & $p_i q_i$ & $p_i$ & $q_i$ \\
\cline{2-6}
\multirow{4}{*}{$\psi_i$} &
$q_i p_i$ & $q_i p_i$ & $0$ & $0$ & $q_i$ \\
& $p_i q_i$ & $0$ & $p_i q_i$ & $p_i$ & $0$ \\
& $p_i$ & $p_i$ & $0$ & $0$ & $p_i q_i$ \\
& $q_i$ & $0$ & $q_i$ & $q_i p_i$ & $0$ \\
\cline{2-6}
\end{tabular}
\caption{Clifford products of EFB elements $\psi_i$ and $\phi_i$ of $\Cl(1,1)$}
\end{table}


We now define the EFB of $\Cl(m, m)$ to be given by all possible sequences
$$
\psi_1 \psi_2 \cdots \psi_m := \Psi \qquad \psi_i \in \{ q_i p_i, p_i q_i, p_i, q_i \} \qquad i = 1,\ldots,m
$$
and since every component $\psi_i$ has just $4$ possible values the basis contains $4^m = 2^{2 m}$ elements (we will reserve Greek capital letters to EFB elements). It's immediate to transform a basis element of the standard $\gamma$ basis, e.g.\ $\gamma_i, \gamma_j, \ldots, \gamma_k$, to a superposition of $2^m$ EFB elements substituting:
\begin{itemize}
\item to each $\gamma_{2 l - 1}$ or $\gamma_{2 l}$ the appropriate sum $(p_l \pm q_l)$ obtainable from (\ref{formula_Witt_basis}),
\item if $\gamma_{2l - 1} \gamma_{2 l}$ are not in $\gamma_i, \gamma_j, \ldots, \gamma_k$ their place is taken by $\anticomm{q_l}{p_l} = \Identity$
\end{itemize}
so that, for example,
$$
\gamma_{1} \gamma_{2 l} = (p_1 + q_1) \anticomm{q_2}{p_2} \anticomm{q_3}{p_3} \cdots (p_l - q_l) \cdots \anticomm{q_m}{p_m}
$$
and the product expands in a sum of precisely $2^m$ EFB elements. Viceversa with (\ref{formula_Witt_basis}) every EFB element can be transformed in a linear superposition of exactly $2^m$ multivectors; these properties reflect the form of orthogonal transformation matrix defined in \cite{BudinichM_2009}.

\smallskip

This basis simplifies the Clifford product of $2$ EFB elements $\Psi$ and $\Phi$ referring them to $\Cl(1,1)$: from (\ref{formula_Witt_basis_properties}) we derive $\psi_i \phi_j = \pm \phi_j \psi_i$ for $i \neq j$ so
$$
\Psi \Phi = \psi_1 \psi_2 \cdots \psi_m \; \phi_1 \phi_2 \cdots \phi_m = \pm \psi_1 \phi_1 \psi_2 \phi_2 \cdots \psi_m \phi_m
$$
and the only relevant products are thus $\psi_i \phi_i$ whose results appear in table~\ref{piqi_table}.

\smallskip

The main characteristics of EFB is that all its elements are simple (also: pure) spinors. We just remind that spinors are minimal left ideals of Clifford algebra and that they are isomorphic to Totally Null Planes (TNP, also: isotropic planes) \cite{BudinichP_1989}. For each spinor $\omega$ we define its corresponding TNP as:
$$
M(\omega) := \{v \in V : v \omega = 0 \mbox{ and } \anticomm{v_a}{v_b} = 0 \quad \forall v_a, v_b \in M(\omega)\}
$$
and the spinor is simple iff the TNP is maximal, i.e. iff $|M(\omega)| = m$.
\begin{MS_Proposition}
\label{EFB_spinors}
The $2^{2 m}$ elements of EFB are simple spinors.
\end{MS_Proposition}
\begin{proof}
We show first that all EFB elements are Weyl spinors, i.e.\ defining the volume element $\Gamma := \gamma_1 \gamma_2 \cdots \gamma_{2 m}$, that any $\Psi = \psi_1 \psi_2 \cdots \psi_m$ is an eigenvector:
\begin{equation}
\label{EFB_are_Weyl}
\Gamma \; \Psi = \h_eigen \; \Psi \qquad \h_eigen = \pm 1
\end{equation}
where we call helicity%
\footnote{chirality could appear more appropriate but helicity is adopted to follow \cite{BudinichP_1989}}%
{} the eigenvalue $\h_eigen$. We first note that $\gamma_{2 i - 1} \gamma_{2 i} = q_i p_i - p_i q_i := \comm{q_i}{p_i}$ and thus $\Gamma = \comm{q_1}{p_1} \comm{q_2}{p_2} \cdots \comm{q_m}{p_m}$. Then we find that for $i \ne j$ $\comm{q_i}{p_i} \psi_j = \psi_j \comm{q_i}{p_i}$ and consequently, that only the products $\comm{q_i}{p_i} \psi_i$ are relevant. With table~\ref{piqi_table} one easily finds
\begin{equation}
\label{commutator_property}
\comm{q_i}{p_i} \psi_i = h_i \psi_i \qquad h_i = \left\{
\begin{array}{l l}
+1 & \quad \mbox{iff $\psi_i = q_i p_i \; \mbox{or} \;  q_i$} \\
-1 & \quad \mbox{iff $\psi_i = p_i q_i \; \mbox{or} \;  p_i$}
\end{array} \right.
\end{equation}
that proves (\ref{EFB_are_Weyl}). The value of $h_i$ depends only on the first null vector appearing in $\psi_i$ and each EFB element has thus also an ``$h-$signature'' that is a vector $(h_1, h_2, \ldots, h_m) \in \{ \pm 1 \}^m$ and clearly $\h_eigen = \prod_{i = 1}^m h_i$.

To prove now that any of these Weyl spinors is simple it is sufficient to show that its associated TNP is maximal, i.e.\ of dimension $m$. For any $\Psi = \psi_1 \psi_2 \cdots \psi_m$ let's call $x_i$ the first null vector appearing in $\psi_i$ then $\my_span{ x_1, x_2, \ldots, x_m}$ is a TNP of maximal dimension $m$ and for any $v \in \my_span{ x_1, x_2, \ldots, x_m}$ we have $v \Psi = 0$, thus it's a simple spinor.
\end{proof}

\smallskip

The ``$g-$signature'' of an EFB element is the vector $(g_1, g_2, \ldots, g_m) \in \{ \pm 1 \}^m$ where $g_i$ is the parity of $\psi_i$ under the main algebra automorphism $\gamma_i \rightarrow - \gamma_i$. With this definition we can easily derive from table~\ref{piqi_table} that
$$
\psi_i \comm{q_i}{p_i} = g_i \comm{q_i}{p_i} \psi_i
$$
and with (\ref{commutator_property}) it follows that for each component $\psi_i \comm{q_i}{p_i} = h_i g_i \psi_i$ and thus for the entire EFB element we have
\begin{equation}
\label{EFB_are_left_Weyl}
\Psi \; \Gamma = \h_eigen \g_eigen \; \Psi \qquad \h_eigen \g_eigen = \pm 1
\end{equation}
where the eigenvalue $\h_eigen \g_eigen$ is composed by the helicity and by $\g_eigen := \prod_{i = 1}^m g_i$, the global parity of the EFB element under the main algebra automorphism. We can resume saying that all EFB elements are not only Weyl eigenvectors, i.e. right eigenvectors of $\Gamma$, but also its left eigenvectors with respective eigenvalues $\h_eigen$ and $\h_eigen \g_eigen = \prod_{i = 1}^m h_i g_i$.

\smallskip

One easily sees that any EFB element $\Psi = \psi_1 \psi_2 \cdots \psi_m$ is uniquely identified by its $h-$ and $g-$signatures: $h_i$ determines the first null vector ($q_i$ or $p_i$) appearing in $\psi_i$ and $g_i$ determines if $\psi_i$ is even or odd. Beyond that $h-$ and $g-$signatures identify also subspaces of the Clifford algebra:
\begin{MS_Proposition}
\label{Clifford_direct_sums}
The Clifford algebra $\Cl(m,m)$, as a vectorial space, is the direct sum of its $2^m$ subspaces of different $h-$signatures
\begin{equation}
\label{Clifford_EFB_direct_sum}
\Cl(m,m) = H_{-\ldots--} \oplus H_{-\ldots-+} \oplus \cdots \oplus H_{+\ldots++}
\end{equation}
where
$$
H_{h_1 h_2 \ldots h_m} = \{ \omega \in \Cl(m,m) \mbox{\rm{ and with $h-$signature }} (h_1, h_2, \ldots, h_m) \}
$$
\end{MS_Proposition}
\begin{proof}
Since for any EFB element $\Psi$ its $h-$signature is defined by
$$
\comm{q_i}{p_i} \Psi = h_i \Psi \qquad i = 1, \ldots, m
$$
it's trivial to see that the span of the $2^m$ EFB elements with same $h-$signature $(h_1, h_2, \ldots, h_m)$ form one subspace and that these $2^m$ subspaces sum up to the whole $\Cl(m,m)$.
\end{proof}

\begin{MS_Corollary}
Identical propositions hold for both $g-$ and $h \circ g-$signatures ($h \circ g$ is the Hadamard (entrywise) product of $h-$ and $g-$signatures vectors).
\end{MS_Corollary}

Observing that in table~\ref{piqi_table} there are $8$ zeros out of $16$ possible products, one can prove easily that only $2^{3 m}$ out of the possible $2^{4 m}$ products of EFB elements are non zero or, more precisely,
\begin{MS_Proposition}
\label{EFB_product}
The Clifford product of two EFB elements $\Psi$ and $\Phi$ is not zero if, and only if,
$$
h_\Psi \circ g_\Psi = h_\Phi
$$
and then the result is an EFB element with $h-$ and $g-$signatures given by
$$
h_{\Psi \Phi} = h_\Psi \qquad \mbox{\rm{and}} \qquad g_{\Psi \Phi} = g_\Psi \circ g_\Phi
$$
\end{MS_Proposition}
\begin{proof}
The proof is simple to do for $\Cl(1,1)$ (see table~\ref{piqi_table}) and it thus applies to each $\psi_i \phi_i$ component, from this derives the desired property.
\end{proof}

\smallskip

We conclude observing that in $\Cl(m,m)$ the standard $\gamma$ basis and EFB have complementary properties. On one side in $\gamma$ basis the algebra can be seen as a direct sum of its $m+1$ grades (\ref{Clifford_direct_sum}) and all products of its basis elements are non zero. On the other hand in EFB the algebra can be seen as a direct sum of $2^m$ subspaces of different signatures (\ref{Clifford_EFB_direct_sum}) while the overwhelming majority of products of EFB elements is zero (only $1$ of $2^m$ is non zero). In addition in EFB spinors have simple expressions whereas vectors have intricate ones.

\section{Matrix isomorphism}
\label{EFB_matrix_isomorphism}
An advantage of this basis is that it maps neatly to the $2^m \times 2^m$ matrices of the algebra $F( 2^m )$ of the representation $\Cl(m,m) \rightarrow \End_F S$. Given $\omega,\phi,\psi \in \Cl(m,m)$ such that $\psi = \omega \phi$ let $f: \Cl(m,m) \rightarrow F( 2^m )$ be the isomorphism of algebras such that
\begin{equation}
\label{isomorphims_property}
C = f(\psi) = f(\omega \phi) = f(\omega) f(\phi) = A B \quad \left\{
\begin{array}{l l}
\omega,\phi,\psi \in \Cl(m,m) \\
A,B,C \in F( 2^m )
\end{array} \right.
\end{equation}
and let's examine first the simple case $m = 1$. $\Cl(1,1)$ has dimension $4$ and it's a simple exercise to verify that the calculation of (here $a_{ij}, b_{kl} \in F$)
$$
\omega \phi = (a_{11} \, q_1 p_1 + a_{12} \, q_1 + a_{21} \, p_1 + a_{22} \, p_1 q_1)(b_{11} \, q_1 p_1 + b_{12} \, q_1 + b_{21} \, p_1 + b_{22} \, p_1 q_1)
$$
establishes the isomorphism of algebras $\Cl(1,1) \cong F( 2 )$ with the map
$$
\omega = a_{11} \, q_1 p_1 + a_{12} \, q_1 + a_{21} \, p_1 + a_{22} \, p_1 q_1 \quad \rightarrow \quad A = \left(\begin{array}{c c} a_{11} & a_{12} \\ a_{21} & a_{22} \end{array}\right) \; \textrm{.}
$$
EFB elements form also a basis in $F( 2 )$ (seen as a vectorial space) and one can more easily verify the isomorphism, stretching a bit the notation, writing
$$
A = \left(\begin{array}{l l} a_{11} \, q_1 p_1 & a_{12} \, q_1 \\ a_{21} \, p_1 & a_{22} \, p_1 q_1 \end{array}\right)
$$
and verifying that the calculation, with usual matrix multiplication rules, satisfies (\ref{isomorphims_property}). To alleviate the notation from now on we omit the field coefficients $a_{ij}$.

\begin{MS_Proposition}
\label{EFB_representation}
for $\Cl(m,m)$ elements expressed in EFB the isomorphism of algebras $\Cl(m,m) \cong F( 2^m )$ is realized by $f: \Cl(m,m) \rightarrow F( 2^m )$ given by
$$
f(\omega) = A_m
$$
and in $A_m$ every entry corresponds to precisely one EFB element so that EFB constitutes also a natural basis in the vectorial space $F( 2^m )$. The matrix $A_m$ is defined recursively by:
\begin{eqnarray*}
A_m & = & A_1 \stackrel{.}{\otimes} A_{m-1} = \left(\begin{array}{r r} q_1 p_1 & q_1 \\ p_1 & p_1 q_1 \end{array}\right) \stackrel{.}{\otimes} A_{m-1} \\
& := & \left(\begin{array}{r r} q_1 p_1 \, A_{m-1} & q_1 \, \Gamma_{m - 1} A_{m-1} \\ p_1 \, \Gamma_{m - 1} A_{m-1} & p_1 q_1 \, A_{m-1} \end{array}\right) \; \rm{.}
\end{eqnarray*}
\end{MS_Proposition}
\begin{proof}
We proceed by induction: we have already seen that the proposition is true for $m = 1$; let's now suppose it true for $m-1$ i.e. that $A_{m-1}$ satisfies (\ref{isomorphims_property}). We note that in the block matrix $A_m$, submatrices $A_{m-1}$ and $\Gamma_{m - 1}$ contain only EFB components in the range $2, \ldots, m$. So, since $q_1$ and $p_1$ don't appear in $A_{m-1}$ it follows (with improper but simple notation)
\begin{eqnarray*}
q_1 p_1 \, A_{m-1} = A_{m-1} \, q_1 p_1 \qquad & q_1 \, A_{m-1} = A^*_{m-1} \, q_1 \\
q_1 p_1 \, \Gamma_{m - 1} = \Gamma_{m - 1} \, q_1 p_1 \qquad & q_1 \, \Gamma_{m - 1} = \Gamma_{m - 1} \, q_1
\end{eqnarray*}
and identical properties for $p_1 q_1$ and $p_1$; here $A^*_{m-1}$ is the matrix where each element has the sign given by its global parity $\theta$, e.g.
$$
A^*_{1} = \left(\begin{array}{r r} q_1 p_1 & - q_1 \\ - p_1 & p_1 q_1 \end{array}\right)
$$
and since for any EFB element $\Gamma \Psi \Gamma = \theta \Psi$ it easily follows that for any $m$
$$
A^*_m = \Gamma_m A_m \Gamma_m \qquad \mbox{and} \qquad A_m = \Gamma_m A^*_m \Gamma_m
$$
since $\Gamma^2 = \Identity$.
We now verify that $A_m$ satisfies (\ref{isomorphims_property}) where, to simplify the following calculations, we drop unnecessary $m-1$ indices and add position indexes to sub matrices $A_{m-1}$:
\begin{align*}
A_m B_m & = \left(\begin{array}{r r} q_1 p_1 \, A_{m-1} & q_1 \, \Gamma_{m - 1} A_{m-1} \\ p_1 \, \Gamma_{m - 1} A_{m-1} & p_1 q_1 \, A_{m-1} \end{array}\right) \left(\begin{array}{r r} q_1 p_1 \, B_{m-1} & q_1 \, \Gamma_{m - 1} B_{m-1} \\ p_1 \, \Gamma_{m - 1} B_{m-1} & p_1 q_1 \, B_{m-1} \end{array}\right) \\
& := \left(\begin{array}{r r} q p \, A_{11} & q \, \Gamma A_{12} \\ p \, \Gamma A_{21} & p q \, A_{22} \end{array}\right) \left(\begin{array}{r r} q p \, B_{11} & q \, \Gamma B_{12} \\ p \, \Gamma B_{21} & p q \, B_{22} \end{array}\right) = \\
& = \left(\begin{array}{l l} q p \, A_{11} \, q p \, B_{11} + q \, \Gamma A_{12} \, p \, \Gamma B_{21} &
		q p \, A_{11} \, q \, \Gamma B_{12} + q \, \Gamma A_{12} \, p q \, B_{22} \\
		p \, \Gamma A_{21} \, q p \, B_{11} + p q \, A_{22} \, p \, \Gamma B_{21} &
		p \, \Gamma A_{21} \, q \, \Gamma B_{12} + p q \, A_{22} \, p q \, B_{22} \end{array}\right) = \\
& = \left(\begin{array}{l l} q p \, (A_{11} B_{11} + \Gamma A_{12}^* \Gamma B_{21}) &
		q \, (\Gamma \Gamma A_{11}^* \Gamma B_{12} + \Gamma A_{12} B_{22}) \\
		p \, (\Gamma A_{21} B_{11} + \Gamma \Gamma A_{22}^* \Gamma B_{21}) &
		p q \, (\Gamma A_{21}^* \Gamma B_{12} +A_{22} B_{22}) \end{array}\right) = \\
& = \left(\begin{array}{l l} q p \, (A_{11} B_{11} + A_{12} B_{21}) &
		q \, \Gamma (A_{11} B_{12} + A_{12} B_{22}) \\
		p \, \Gamma (A_{21} B_{11} + A_{22} B_{21}) &
		p q \, (A_{21} B_{12} +A_{22} B_{22}) \end{array}\right) = \\
& = \left(\begin{array}{l l} q p \, C_{11} &
		q \, \Gamma C_{12} \\
		p \, \Gamma C_{21} &
		p q \, C_{22} \end{array}\right) = C_m \tag*\qedhere
\end{align*}
\end{proof}

A simple corollary gives the recursive definition of the volume element:
\begin{equation}
\label{Gamma_recursive}
\Gamma_m = \Gamma_1 \otimes \Gamma_{m-1} = \left(\begin{array}{r r} 1 & 0 \\ 0 & -1 \end{array}\right) \otimes \Gamma_{m-1} \; \rm{.}
\end{equation}

A more interesting result is:
\begin{MS_Corollary}
\label{columns_are_MLI}
The columns of $A_m$ are minimal left ideals of $\Cl(m, m)$ and are formed by EFB elements with the same $h \circ g-$signature.
\end{MS_Corollary}
\begin{proof}
We start observing that the EFB elements in the rows of the matrix have all identical $h-$signatures how it is clear from $A_1$ and from the recursive construction of $A_m$. As a consequence the EFB elements in each column contain all $2^m$ possible $h-$signatures. Moreover in each column the termwise product of $h-$ and $g-$signatures is constant throughout the column. Also this can be proved easily from the recursive construction of $A_m$, for example the rightmost column has for each component $h_i g_i = -1$ and in it we can recognize (forgetting the irrelevant sign) the usual Fock basis of spinor space \cite{BudinichP_1989}.

We prove now that the Clifford product of any couple of EFB elements is either $0$ or one of the elements of the column containing the second term of the product thus proving that the elements of this column form a left ideal. To prove this it's sufficient to prove that $h \circ g-$signature of an EFB element is invariant by left multiplication from another EFB element. This is clear since, by proposition~\ref{EFB_product} EFB elements product $\Psi \Phi$ is not zero only if $h_\Psi \circ g_\Psi = h_\Phi$ and the result has $h-$ and $g-$signatures given respectively by $h_\Psi$ and $g_\Psi \circ g_\Phi$. For the non zero result we thus have
$$
h_{\Psi \Phi} \circ g_{\Psi \Phi} = h_\Psi \circ g_\Psi \circ g_\Phi = h_\Phi \circ g_\Phi
$$
and thus the $h-$ and $g-$signatures product of $\Phi$ is invariant.
\end{proof}

\smallskip

For example the isomorphic matrix of $\Cl(2,2)$ with $h$ (rows) and $h \circ g$ (columns) signatures is:
$$
A_2 = \bordermatrix{& ++ & +- & -+ & -- \cr
++ & q_1 p_1 \, q_2 p_2 & q_1 p_1 \, q_2 & q_1 \, q_2 p_2 & q_1 \, q_2 \cr
+- & q_1 p_1 \, p_2 & q_1 p_1 \, p_2 q_2 & - q_1 \, p_2 & - q_1 \, p_2 q_2 \cr
-+ & p_1 \, q_2 p_2 & p_1 \, q_2 & p_1 q_1 \, q_2 p_2 & p_1 q_1 \, q_2 \cr
-- & - p_1 \, p_2 & - p_1 \, p_2 q_2 & p_1 q_1 \, p_2 & p_1 q_1 \, p_2 q_2 \cr }
$$
and we will call the rightmost column the {\em standard} Fock basis of spinor space $S_F$ i.e.
\begin{equation}
\label{standard_Fock_basis}
S_F := \{ \Omega \in \textrm{EFB} : h_\Omega \circ g_\Omega = \{-1\}^{2^m} \} \; \textrm{.}
\end{equation}

\smallskip

As a final remark we observe that this isomorphism provides the provably faster algorithm for actual Clifford product evaluations \cite{BudinichM_2009} and results a factor $2^m$ faster than usual algorithms based on $\gamma$ matrices.

\section{Multiple spinor spaces}
\label{multiple_spinor_spaces}
Propositions~\ref{Clifford_direct_sums} and \ref{columns_are_MLI} show that $\Cl(m, m)$, as a vectorial space, is the direct sum of subspaces of different $h \circ g-$signatures that are also minimal left ideals of $\Cl(m, m)$ and thus spinor spaces $S_{h \circ g}$. Moreover they correspond to different columns of the isomorphic algebra of $F( 2^m )$.

All the EFB elements of one of these subspaces form a base of their spinor space $S_{h \circ g}$ and are also left eigenvectors of $\Gamma$ of eigenvalue $\h_eigen \g_eigen$ (\ref{EFB_are_left_Weyl}). So speaking of a spinor space $S$ it is always necessary to specify its $h \circ g-$signature. We clarify this with an example: it is known \cite{BudinichP_1989} that maximal TNP are isomorphic to simple spinors of $S$ but this correspondence is obscured if we don't specify one spinor space. For example in $\Cl(2, 2)$ the $4$ EFB elements $p_1 p_2, p_1 q_1 p_2, p_1 p_2 q_2$ and $p_1 q_1 p_2 q_2$ are all simple spinors, are linearly independent and all have the same TNP, namely $\my_span{p_1,p_2}$. Specifying the $h \circ g-$signature, for example choosing the $h \circ g = \{-1\}^{2^m}$ of $S_F$, we have the only simple spinor $p_1 q_1 p_2 q_2$ and the isomorphism is reestablished.

In general the $h-$signature of an EFB element $\Psi$ fixes uniquely the associated maximal TNP $M(\Psi)$ and there are $2^m$ EFB elements with same $h-$signature and all possible $2^m g-$signatures. These EFB elements form one of the subspaces $H$ of proposition~\ref{Clifford_direct_sums} and they can be obtained from $\Psi$ replacing every $\psi_i$ with its counterpart with same first null vector and opposite $g-$signature, i.e.\ $p_i \leftrightarrow p_i q_i$ and $q_i \leftrightarrow q_i p_i$. So for each TNP we have $2^m$ different, linearly independent simple spinors such that for any of them (and even for any of their linear combinations) $v \Psi = 0$.

It is simple to see that all the EFB elements of one of these subspaces $H$ can be obtained by one of them $\Psi$ right multiplying it by the unit vectors $(p_i + q_i)$ that has the effect of flipping just $g_i$ in the EFB element $\Psi$. Since the Pin group consists of products of unit vectors (and its subgroup Spin consists of products of even sequences of unit vectors) the action of its elements generate the entire subspace $H$ (while the action of Spin generate all EFB elements with same eigenvalue $\h_eigen \g_eigen$).

This subject certainly deserves deeper investigations also in view that multiple spin spaces $S_{h \circ g}$ have been proposed for mirror particles \cite{Pavsic_2010} and one should thus explore the possible physical implications of (\ref{EFB_are_left_Weyl}).

\section{Properties of simple spinors in EFB}
\label{spinor_properties_in_EFB}

The linear superposition of 2 EFB elements of the same spinor space $S$ can be a simple spinor (unless explicitly specified we will refer here to the spinor space of the standard Fock basis $S_F$ (\ref{standard_Fock_basis})):
\begin{MS_Proposition}
\label{EFB_sum}
Let $\Omega$ and $\Phi$ be different EFB elements of the same spinor space $S$ of $\Cl(m, m)$ then a linear combination $a \Omega + b \Phi$ ($a, b \in F$) is simple if, and only if, the size of the intersection of their TNP is $| M(\Omega) \cap M(\Phi)| = m-2$ and $h-$ and $g-$signatures of $\Omega$ and $\Phi$ are equal in the $m-2$ EFB components with same $h-$signature and opposite in the remaining 2.%
\footnote{This is a slight extension of proposition~$5$ of \cite{BudinichP_1989} and III$.1.12$ of \cite{Chevalley_1954}, we give here a different proof based on more elementary arguments. We remark that without the hypothesis of same spinor space $S$ the proposition does not hold and one can build linear combinations of up to $2^m$ EFB elements of $\Cl(m, m)$ that annihilate a maximal TNP. It's easy to adapt this and the following propositions to be valid in the whole $\Cl(m, m)$ seen as a vectorial space.
}%
\end{MS_Proposition}
\begin{proof}
Since $| M(\Omega) \cap M(\Phi)| = m-2$ without loss of generality we may assume $M(\Omega) = \my_span{q_1, q_2, q_3, \ldots,q_m}$ and $M(\Phi) = \my_span{p_1, p_2, q_3,\ldots,q_m}$. For any vector $u \in M(\Omega) \cap M(\Phi) = \my_span{q_3,\ldots,q_m}$ obviously $u \Omega = u \Phi = 0$ and for them trivially $u (a \Omega + b \Phi) = 0$ so, to prove that $a \Omega + b \Phi$ is simple, we need $2$ more linearly independent null vectors to form a maximal TNP.

We show now that a vector $u$ such that $u (a \Omega + b \Phi) = 0$ cannot be $u \in \overline{M(\Omega) \cup M(\Phi)} = \my_span{p_3,\ldots,p_m}$ because this would imply
$$
a u \Omega = -b u \Phi \ne 0
$$
that in turn implies with proposition~\ref{EFB_product} that $\Omega$ and $\Phi$ necessarily have identical $h-$ and $g-$signatures against the hypothesis of their difference.

So to satisfy $u (a \Omega + b \Phi) = 0$ the only possibility is that $u \in M(\Omega) \cup M(\Phi) - M(\Omega) \cap M(\Phi)$ that in our case is $u \in \my_span{p_1,p_2,q_1,q_2}$ and substituting this form in the relation to be satisfied we get ($c_i, d_i \in F$)
$$
(c_1 p_1 + c_2 p_2 + d_1 q_1 + d_2 q_2) (a \Omega + b \Phi) = 0
$$
that becomes
\begin{equation}
\label{simple_spinor_sum_equation}
a (c_1 p_1 + c_2 p_2) \Omega = - b(d_1 q_1 + d_2 q_2) \Phi \ne 0
\end{equation}
in which both terms are non zero by hypothesis. The parts of $\Omega$ and $\Phi$ unaffected by left multiplication by the vectors (i.e. in our case EFB components $3,\ldots,m$) must necessarily be identical to satisfy this equation. We may thus concentrate on the first $2$ EFB components of $\Omega$ and $\Phi$, respectively $\omega_1, \omega_2$ and $\phi_1, \phi_2$, thus the reduced relation to be satisfied is:
$$
a (c_1 \omega_1' \omega_2 + g_{\omega_1} c_2 \omega_1 \omega_2') = - b(d_1 \phi _1' \phi _2 + g_{\phi_1} d_2 \phi_1 \phi _2')
$$
where the primed components indicate the initial component left multiplied by the corresponding vector. Given EFB properties it is clear that the only possibility to satisfy this equality is to have separately
$$
a c_1 \omega_1' \omega_2 = - g_{\phi_1} b d_2 \phi_1 \phi _2' \qquad g_{\omega_1} a c_2 \omega_1 \omega_2' = - b d_1 \phi _1' \phi _2
$$
that to be satisfied imply for the EFB components
\begin{align*}
h_{\omega_1} = h_{q_1} = 1 & \qquad h_{\omega_2} = h_{q_2} = 1 \\
h_{\phi_1} = h_{p_1} = -1 & \qquad h_{\phi_2} = h_{p_2} = -1 \\
g_{\omega_1} = - g_{\phi_1} & \qquad g_{\omega_2} = - g_{\phi_2}
\end{align*}
and for the field coefficients
$$
a c_1 = - g_{\phi_1} b d_2 \qquad g_{\omega_1} a c_2 = - b d_1
$$
Supposing that $a, b \ne 0$ (the other cases are trivial) it follows that the vectors
$$
u \in \my_span{p_1 - g_{\phi_1} \frac{a}{b} q_2, \: g_{\omega_1} p_2 - \frac{a}{b} q_1}
$$
span a 2-dimensional space, are null and annihilate spinor $a \Omega + b \Phi$ that is thus simple. The conditions that $\Omega$ and $\Phi$ have to satisfy are
\begin{eqnarray*} 
\mbox{for } i = 1,2 & h_{\omega_i} = - h_{\phi_i} & g_{\omega_i} = - g_{\phi_i} \\
\mbox{for } i = 3,\ldots,m & h_{\omega_i} = h_{\phi_i} & g_{\omega_i} = g_{\phi_i}
\end{eqnarray*}

We conclude showing that if $| M(\Omega) \cap M(\Phi)| \ne m-2$ then $a \Omega + b \Phi$ cannot be simple. Any simple spinor is necessarily a $\Gamma$ eigenvector and this holds true also for EFB elements that have eigenvalue $\h_eigen = \prod_{i = 1}^m h_i$. If $\Omega + \Phi$ is a simple spinor it must be a $\Gamma$ eigenvector and thus $\prod_{i = 1}^m h_{\omega_i} = \prod_{i = 1}^m h_{\phi_i}$. Since $h-$signature defines uniquely the TNP associated to EFB elements it follows that $| M(\Omega) \cap M(\Phi)| = {m - 2 k}$ with $0 < 2k \le m$.

Supposing e.g. $| M(\Omega) \cap M(\Phi)| = m-4$ the equation corresponding to (\ref{simple_spinor_sum_equation}) would be now
$$
a (c_1 p_1 + c_2 p_2 + c_3 p_3 + c_4 p_4) \Omega = - b(d_1 q_1 + d_2 q_2 + d_3 q_3 + d_4 q_4) \Phi
$$
and it is obvious that this can never be satisfied (consider for example $\Omega = q_1 q_2 q_3 q_4 \ldots$ and $\Phi = p_1 p_2 p_3 p_4 \ldots$ then $p_i \Omega$ can never be equal to any $q_i \Phi$) and in general is satisfied only when the intersection is of size $m-2$ because only in this case multiplication by a vector may change one component of $\Omega$ into another of $\Phi$.
\end{proof}

This result easily generalizes to generic simple spinors:
\begin{MS_Corollary}
\label{EFB_sum_corollary_1_generic_spinors}
Let $\omega$ and $\phi$ be linearly independent simple spinors of the same spinor space $S$ of $\Cl(m, m)$ then a linear combination $a \omega + b \phi$ ($a, b \in F$) is simple if, and only if, the size of the intersection of their TNP is $| M(\omega) \cap M(\phi) | = m-2$.
\end{MS_Corollary}
\begin{proof}
To prove this we apply proposition~2 of \cite{BudinichP_1989} that, easily extended 
to any number of spinors and rephrased in EFB jargon, asserts: ``given $2$ or more linearly independent simple spinors then there exists a basis (\ref{formula_Witt_basis}) such that these spinors are different EFB elements of the same spinor space'' that sends us back to the previous case.
\end{proof}

\begin{MS_Proposition}
\label{EFB_sum_many}
Given $k$ linearly independent simple spinors of the same spinor space $S$ of $\Cl(m, m)$ such that, for any two of them, the size of the intersection of their respective TNP's is $m-2$, then all of their linear combinations are simple and the following bounds hold:
\begin{align*}
k & \le m & \mbox{\rm{with the exception of}}\\
k & \le m+1 & \mbox{\rm{for}} \quad m = 3
\end{align*}

\end{MS_Proposition}
\begin{proof}
We start proving a reduced version, namely for $3$ EFB elements and, without loss of generality, we take this example as a guide in our reasoning:
\begin{eqnarray*} 
\Omega = q_1 q_2 q_3 q_4 \cdots q_m & \quad M(\Omega) = \my_span{q_1, q_2, q_3, q_4, \ldots,q_m} \\
\Phi = p_1 q_1 p_2 q_2 q_3 q_4 \cdots q_m & \quad M(\Phi) = \my_span{p_1, p_2, q_3, q_4, \ldots,q_m} \\
\Psi = p_1 q_1 q_2 p_3 q_3 q_4 \cdots q_m & \quad M(\Psi) = \my_span{p_1, q_2, p_3, q_4, \ldots,q_m}
\end{eqnarray*}
By proposition~\ref{EFB_sum} we already know that $\Omega - \Phi$ is simple with $M(\Omega - \Phi) = \my_span{q_1 - p_2, q_2 + p_1, q_3, q_4, \ldots,q_m}$; we now show that from the hypothesis $| M(\Psi) \cap M(\Omega) | = | M(\Psi) \cap M(\Phi) | = m-2$ descends $| M(\Psi) \cap M(\Omega - \Phi) | = m - 2$ so that we can apply corollary~\ref{EFB_sum_corollary_1_generic_spinors} to $\Omega - \Phi$ and $\Psi$ completing the proof of the reduced part of this proposition.

First of all we remind that each of the $2^m$ TNP's is obtained choosing among $p_i$ and $q_i$ for its $i$-th basis vector. Let's consider the $2$ vectors of $M(\Omega) - M(\Omega) \cap M(\Phi)$, in our example $q_1$ and $q_2$. We first show that it is impossible that both vectors are also in $M(\Psi) \cap M(\Omega)$, by hypothesis of dimension $m - 2$, because this would imply that $M(\Psi)$ contains two other vectors that are not in $M(\Omega)$, let us say $p_r$ and $p_s$, but this would in turn violate the hypothesis $| M(\Psi) \cap M(\Phi) | = m - 2$. Similarly, if we suppose that neither $q_1$ nor $q_2$ are in $M(\Psi) \cap M(\Omega)$, it follows that $p_1$ and $p_2$ must be in $M(\Psi)$ and this would give $| M(\Psi) \cap M(\Phi) | = m$ against our initial hypothesis. We must thus conclude that one, and only one, of $q_1$ and $q_2$ must belong to $M(\Psi) \cap M(\Omega)$. With the same reasoning we can prove that one, and only one, of the vectors of $M(\Phi) - M(\Omega) \cap M(\Phi)$, in our example $p_1$ and $p_2$, must belong to $M(\Psi) \cap M(\Phi)$. This also easily proves that necessarily $| M(\Psi) \cap M(\Omega) \cap M(\Phi) | = m - 3$, that in our example are the vectors $q_4, \ldots, q_m$.

We observe now that if $q_1 \in M(\Psi) \cap M(\Omega)$ than $p_1$ cannot belong to $M(\Psi)$ since $p_i$ and $q_i$ can never be in the same TNP. It follows that $M(\Psi)$ contains either $q_1$ and $p_2$ or $p_1$ and $q_2$; in either case $M(\Psi)$ has one direction in common with $M(\Omega - \Phi)$ to be added to the $m - 3$ directions that are in common with $M(\Psi) \cap M(\Omega) \cap M(\Phi)$ and also with $M(\Omega - \Phi)$. So we have shown that $| M(\Psi) \cap M(\Omega - \Phi) | = m - 2$ and we can apply corollary~\ref{EFB_sum_corollary_1_generic_spinors} to simple spinors $\Omega - \Phi$ and $\Psi$ to conclude that any linear combination of $\Omega$, $\Phi$ and $\Psi$ is a simple spinor. Given the arbitrary choice of the $3$ EFB elements this concludes the proof of the reduced version of the proposition.

We now apply the same basis transformation used in the demonstration of corollary~\ref{EFB_sum_corollary_1_generic_spinors} to show that this proposition holds for $3$ generic simple spinors $\omega_1, \omega_2, \omega_3 \in \Cl(m, m)$. Having proved the case of $k = 3$ it is simple to extend it to any $k$: it suffices to start from two simple spinors adding the remaining spinors one at the time iterating the proof at each step. For example to show that any linear combination of $\omega_1, \omega_2, \omega_3$ and $\omega_4$ is simple it's sufficient to apply our result to the $3$ simple spinors $\omega_1 + \omega_2, \omega_3$ and $\omega_4$ and since we know that the result holds for any linear combination $\omega_1 + \omega_2$ it must hold for all $4$ simple spinors.

\smallskip

To prove the upper bounds on $k$ we start observing that it's sufficient to prove it for EFB since any set of $k$ linearly independent simple spinors can be transformed in $k$ EFB elements. For $m = 1$ there are just $2$ TNP's of dimension $1$ whereas for $m = 2$ there are only $2$ TNP that have size of the intersection $m - 2 = 0$, namely $p_1 p_2$ and $q_1 q_2$ so obviously $k \le 2$. For $m > 2$ we prove the bounds proving a related property of binary vectors. First of all we observe that in EFB the set of TNP's is isomorphic to the set of $h-$signatures that are elements of the set $\{ \pm 1 \}^m \subset \R^m$ and thus $2$ TNP's with an intersection of dimension $m-2$ have $h-$signatures such that $h_{(1)} \cdot h_{(2)} = m - 4$. We prove our bound observing that for $m > 3$ there cannot be more than $m$ signature vectors $h_{(i)} \in \{ \pm 1 \}^m$ such that for any two of them $h_{(i)} \cdot h_{(j)} = m - 4$ because these binary vectors are linearly independent in $\R^m$. To prove this let's suppose the contrary and that there are $r$ vectors with the given scalar products such that for given coefficients $a_1, a_2, \ldots, a_r$ (not all $0$) one has
$$
\sum_{i = 1}^r a_i h_{(i)} = 0
$$
scalar multiplying this relation by any of the vectors $h_{(j)}$ and remembering that $h_{(i)} \cdot h_{(j)} = m - 4 (1 - \delta_{ij})$ one easily gets
$$
\frac{4 - m}{4} \sum_{i = 1}^r a_i = a_j
$$
that, given the arbitrary choice of $j$, proves that all coefficients are equal $a_1 = a_2 = \cdots = a_r \ne 0$ and to satisfy the previous relation one needs
$$
\frac{4 - m}{4} r = 1 \quad \rm{.}
$$
Since both $m$ and $r$ are positive integers this relation can be satisfied only for $m \le 3$. For $m > 3$ the relation can never be satisfied that proves that the vectors $h_{(i)}$ are linearly independent and thus there cannot exist more than $m$ of them; this proves the last bound. We are left with the particular case $m = 3$ where previous relation gives $r = 4$ and there are indeed $4$, linearly dependent, vectors $h_{(i)}$, namely $(1,1,1)$, $(1,-1,-1)$, $(-1,1,-1)$ and $(-1,-1,1)$ with all scalar products equal to $m - 4 = -1$ that correspond to $4$ simple spinors $q_1 q_2 q_3$, $q_1 p_2 q_2 p_3 q_3$, $p_1 q_1 q_2 p_3 q_3$ and $p_1 q_1 p_2 q_2 q_3$ with required property and such that any of their linear combinations is simple. The appendix contains an example showing that the bound is strict (i.e. that one can always form a simple spinor with $m$ EFB elements).
\end{proof}

\smallskip

This result provides a different explanation of why, for a generic $\psi \in S$, the request of being a Weyl spinor (i.e. $\Gamma \psi = \pm \psi$) is necessary and sufficient for $\psi$ to be simple for $m \le 3$ becoming only necessary for $m > 3$ \cite{BudinichP_1989}.

It's easy to see that in matrix form the $\psi^{\pm}$ eigenvectors of $\Gamma$ for $m=1$ are respectively:
$$
\Gamma_1 = \left(\begin{array}{r r} 1 & 0 \\ 0 & -1 \end{array}\right) \qquad
\psi^{+}_1 = \left(\begin{array}{c} 1 \\ 0 \end{array}\right) \quad \psi^{-}_1 = \left(\begin{array}{c} 0 \\ 1 \end{array}\right)
$$
and with the $\Gamma$ recursive definition (\ref{Gamma_recursive}) they can be easily found for $m > 1$:
$$
\Gamma_{m} = \left(\begin{array}{c c} \Gamma_{m-1} & 0 \\ 0 & - \Gamma_{m-1} \end{array}\right) \qquad
\psi^{+}_{m} = \left(\begin{array}{c} \psi^{+}_{m-1} \\ \psi^{-}_{m-1} \end{array}\right) \quad \psi^{-}_{m} = \left(\begin{array}{c} \psi^{-}_{m-1} \\ \psi^{+}_{m-1} \end{array}\right)
$$
and it's obvious that each $\psi^{\pm}_{m}$ contains $2^{m - 1}$ zeros. More precisely the span of the $2^{m - 1}$ non zero EFB elements of $\psi^{\pm}_{m} \in S$ form $2^{m - 1}$-dimensional subspaces of definite helicity $S_\pm$ and moreover $S = S_+ \oplus S_-$. The dimensions of these subspaces for $m = 1,2,3$ are respectively $1,2,4$ that match the bound of proposition~\ref{EFB_sum_many}. For $m > 3$ the bound is violated and $\psi \in S_\pm$ are not simple spinors unless they satisfy further conditions, the so called ``constraint equations''. We observe also that for $m \le 3$ simple spinors are subspaces of $S$ while for $m > 3$ they form a manifold containing very many ``totally simple planes'' of dimension $m$ made of simple spinors.

This result give a partial answer to the problem of the constraint equations that a spinor have to satisfy in order to be simple for $m > 3$ since it should be possible to use this property to build explicitly the more general simple spinors for any $m$. This will be the subject of a forthcoming paper.

\newpage

\section{Conclusions}
\label{my_Conclusions}
We have shown that, beyond providing the fastest way to actually calculate Clifford products, EFB offers several advantages:
\begin{itemize}
\item refers many properties of $\Cl(m, m)$ down to $\Cl(1, 1)$,
\item is formed only by simple spinors and has a simple map to the isomorphic matrix algebra,
\item renders explicit the existence of $2^m$ spinor spaces in $\Cl(m, m)$ characterized by being left $\Gamma$ eigenvectors,
\item allows to prove that spinor spaces $S$ contain totally simple planes of dimension $m$ made entirely of simple spinors.
\end{itemize}
About this last point we remark that it is intriguing that the same bound $m$ applies both to the dimensions of a TNP subspace of the vector space $V$ and to a totally simple plane made entirely of simple spinors in $S$ and that the case $m=3$ is exceptional.

\vspace{10 cm}

\subsection*{Acknowledgments}
I warmly acknowledge the many fruitful discussions on Clifford algebras and spinors with my father Paolo that deepen their roots in his last effort \cite{BudinichP_2009}.

\newpage

\appendix
\section*{Appendix: combining $m$ EFB elements to form a simple spinor}
Referring to the example worked out in the proof of proposition~\ref{EFB_sum_many} we may write for the simple spinor
$$
\Omega - \Phi + \Psi = (1 + p_1 p_2 + p_1 p_3) \Omega = [1 + p_1 (p_2 + p_3)] \Omega
$$
to which correspond the TNP
$$
M(\Omega - \Phi + \Psi) = \my_span{q_1 - (p_2 + p_3), \; q_2 + p_1, q_3 + p_1, q_4, \ldots,q_m}
$$
how it's easy to verify. More in general for $m > 3$ the typical set of $k \le m$ simple spinors with reciprocal TNP's intersections of dimension $m - 2$ is obtained by the succession
\begin{multline*}
q_1 q_2 q_3 q_4 \cdots q_m, \quad p_1 q_1 p_2 q_2 q_3 q_4 \cdots q_m, \quad p_1 q_1 q_2 p_3 q_3 q_4 \cdots q_m, \\
\quad p_1 q_1 q_2 q_3 p_4 q_4 \cdots q_m, \quad \ldots, \quad p_1 q_1 q_2 q_3 q_4 \cdots q_{k-1} p_k q_k q_{k+1} \cdots q_m
\end{multline*}
and the simple spinor corresponding to the alternating sum of these terms can be written (as before $\Omega = q_1 q_2 \cdots q_m$)
$$
\left(1 + p_1 \sum_{i = 2}^k p_i \right) \Omega
$$
and defining the null vector $v := \sum_{i = 2}^k p_i$ the corresponding TNP is given by
$$
\my_span{q_1 - v, q_2 + p_1, q_3 + p_1, \ldots, q_{k} + p_1, q_{k+1}, \ldots, q_m}
$$
how is simple to verify: it is fairly obvious that this span forms a maximal TNP. It's simple to show also that
$$
(q_1 - v)(1 + p_1 v) \Omega = (q_1 - v + q_1 p_1 v) \Omega = - v (1 - q_1 p_1) \Omega = - v p_1 q_1 \Omega = 0
$$
and for the other vectors ($ 2 \le j \le k $)
\begin{multline*}
(q_{j} + p_1)(1 + p_1 v) \Omega = (q_j + p_1 - p_1 q_j v) \Omega = [p_1 - p_1(q_j p_j - \sum_{i \ne j \: i = 2}^k p_i q_j )] \Omega = \\
= p_1 (1 - q_j p_j) \Omega = p_1 p_j q_j \Omega = 0 \; \textrm{.}
\end{multline*}
This example shows constructively that the bound of proposition~\ref{EFB_sum_many} is strict.

\newpage

\bibliographystyle{plain} 

\bibliography{mbh}

\opt{final_notes}{
\include{Paper_notes}
} 

\end{document}